\newcommand{\prl}[3]{ Phys.\ Rev.\ Lett.\      {\bf  #1}   (#2)   #3}

\newcommand{\prd}[3]{ Phys.\ Rev.\             {\bf D#1}   (#2)   #3}
\newcommand{\plb}[3]{ Phys.~Lett.\             {\bf B#1}   (#2)   #3}
\newcommand{\npb}[3]{ Nucl.~Phys.\             {\bf B#1}   (#2)   #3}

\newcommand{\pol}[3]{ Acta Phys.\ Polon.\      {\bf  #1}   (#2)   #3}

\newcommand{\for}[3]{ Fortschr.\ Phys.\        {\bf  #3}   (#2)   #3}

\newcommand{\zpc}[3]{ Z.~Phys.\                {\bf C#1}   (#2)   #3}
\newcommand{\GeV}{\mbox{$\mathrm{\ GeV}$}}

\newcommand{\beq}{\begin{equation}}
\newcommand{\eeq}{\end{equation}}
\newcommand{\lsim}{\mathrel{\raise.5ex\hbox{$<$}\kern-.75em\raise
                                             -.5ex\hbox{$\sim$}}}
\newcommand{\gsim}{\mathrel{\raise.5ex\hbox{$>$}\kern-.75em\raise
                                             -.5ex\hbox{$\sim$}}}

\font\elevenit=cmti10 scaled\magstep1
\documentstyle[12pt]{article}
\begin{document}
\begin{titlepage}
\baselineskip=16pt
\begin{flushright}
CERN-TH/95-202\\
BARI-TH/211-95\\
\end{flushright}
\begin{centering}
\vspace{10mm}{\LARGE\bf
                The Higgs Boson Mass \\[2mm]
               from Precision Electroweak Data\\}
\vspace{12mm}%
                                       {\Large  John Ellis~$^a$},
                                       {\Large  G.L. Fogli~$^b$} and
                                       {\Large  E. Lisi~$^{b,c}$}
\\
 \vspace{6mm}%
{\elevenit$^a$ Theory Division, CERN, CH-1211, Geneva, Switzerland
  \\
          $^b$ Dipartimento di Fisica and Sezione INFN di Bari, Bari,
Italy\\
          $^c$ Institute for Advanced Study, Princeton, NJ 08540
  \\}
 \vspace{16mm}%
                                          {\bf Abstract}
  \\
\end{centering}
\vspace{3mm}%
%\begin{quote}
We present a new global fit to precision electroweak data, including
new low- and high-energy data and analyzing the radiative
corrections  arising from the minimal symmetry breaking sectors of
the Standard Model (SM) and its supersymmetric extension (MSSM). It
is shown that present data favor a Higgs mass of ${\cal O}(M_Z)$:
\vspace{-2mm}
$$
 M_H = 76\begin{array}{l}{+152}\\[-6mm]\\
{-\,\,50}\end{array}\quad\GeV\quad.
\vspace{-2mm}
$$
We confront our analysis with (meta)stability and perturbative
bounds on the SM Higgs mass, and the theoretical upper bound on the
MSSM Higgs mass. Present data do not discriminate significantly
between the SM and MSSM Higgs mass ranges. We comment in passing on
the sensitivity of the Higgs mass determination to the values of
$\alpha(M_Z)$ and $\alpha_s(M_Z)$.
%\end{quote}
\vfill
  \begin{flushleft}%  ...................
                      CERN-TH/95-202  \\
                      BARI-TH/211-95  \\
                      July 1995
  \end{flushleft}%    ...................
\end{titlepage}

%-----------------------------------------------------------------------
\newpage
\baselineskip=18pt
\section{Introduction}

The discovery of the top quark by the CDF \cite{CD95} and D0
\cite{D095} collaborations with a  mass that agrees to within 10\%
with that predicted from precision electroweak data
\cite{El94,EW95,Fits} constitutes an impressive success for the
Standard Model, confirming its predictivity at the level of quantum
loops.

The fact that the Standard Model (SM) is renormalizable \cite{Reno}
if and only if the top quark and the Higgs boson are included
implies that loop corrections are sensitive to the masses
$(m_t,\,M_H)$ of these particles, with the sensitivity to $m_t$
being much stronger than that to $M_H$ \cite{Ve77}.

Low-energy data and early measurements of $M_W$ were used to
constrain $m_t$ before the start of SLC and LEP physics
\cite{LE88}, at which time the sensitivities of $Z$ decay
observables to $m_t$ and $M_H$ were well known \cite{ZLEP}. It was
pointed out that these calculations could be used to predict $m_t$
once a precise measurement of $M_Z$ became available \cite{El89}.
Subsequently, the use of this and other precision $Z$ decay
observables to predict $m_t$ has become a major focus of interest
in LEP and SLC physics \cite{El92,El94,EW95,Fits}.

For some time now, the precision electroweak data have also
exhibited some sensitivity to $M_H$ \cite{El90,Ag92}, and the
success of the $m_t$ prediction is now shifting the focus of
interest to the prediction of $M_H$ \cite{El94,EW95,Ch95},
particularly in view of the advent of LEP 2 physics and the drive
to optimize the continuation of the Higgs search at the LHC. The
precision electroweak data have consistently favored $M_H<300$ GeV,
and it is important when considering the maximum energy of LEP 2
and the low-mass Higgs search at the LHC to understand how
seriously to take this trend.

This understanding is also relevant for indications on the
direction of particle physics beyond the Standard Model. The
(meta)stability of the electroweak vacuum \cite{Stab,Meta} imposes
a lower limit on $M_H$ that depends on $m_t$ and the scale
$\Lambda_V$ up to which the Standard Model effective potential is
assumed to represent physics accurately. There is also an upper
limit on $M_H$ that follows from requiring the Standard Model
couplings to remain perturbative up to a scale $\Lambda_P$
\cite{Sherrev}. On the other hand, in the Minimal Supersymmetric
extension of the Standard Model (MSSM), constraints on the form of
the effective potential impose an intrinsic upper limit on the
lightest Higgs mass of order 150 GeV. It is thus important to see
how the indirect determinations of the Higgs mass in the SM and the
MSSM compare with the above limits, the composite (technicolor)
Higgs option being in serious conflict with electroweak data
\cite{El95}.

The main purpose of this paper is to discuss $M_H$ in the light of
the recent direct determination of $m_t$ and the latest round of
precision electroweak data from LEP and elsewhere. We argue that
the combined direct and indirect data now favour significantly
$M_H<300$ GeV. We confront the indirect determination of $M_H$ with
theoretical bounds from vacuum stability and supersymmetry.  We
conclude that all the experimental results and theoretical  bounds
are consistent with both the  SM and the MSSM. We attempt to
quantify the relative probabilities of the Higgs mass ranges in the
SM and MSSM, finding an indication that the MSSM may be preferred.
This indication is not yet significant, but may become so in the
future. In passing, we supplement our discussion of the electroweak
precision data analysis with a more technical issue: the impact of
$\alpha(M_Z)$, in the light of various recent re-evaluations
\cite{Eide,Mart,Burk,Swar}, showing in particular how future
measurements of the muon $g_\mu\!-\!2$ \cite{E821} can improve our
knowledge of $\alpha(M_Z)$.

\section{Data Analysis and Fits to $m_t$}

Our procedure for fitting the available electroweak data is
basically the same as we have described in previous works
\cite{El94,El92,El90}, so here we just comment on the new data that
have recently become available, and the way we treat them.

Foremost are the measurements of $m_t$ by CDF \cite{CD95}:
$m_t=176\pm8\,$(stat)$\;\pm 10\,$(sys) GeV, which we interpret as
$176\pm 13$ GeV, and by D0 \cite{D095}:
$m_t=199^{+19}_{-21}\,$(stat)$\;\pm 22\,$(sys) GeV, which we
interpret as $199\pm 29$ GeV. In 1994, the compatibility of the
indications from CDF \cite{CD94} and the absence of an indication
from D0 \cite{D094} was an issue, as was compatibility with
theoretical calculations of the $t\overline t$ production
cross-section \cite{TTcs}. These are no longer issues, as the CDF
and D0 measurements of $m_t$ are  highly compatible, allowing us to
combine them to obtain $m_t = 181\pm 12$ GeV, and the cross-section
measurements are now also highly compatible with each other and
quite consistent with the theoretical calculations.

In addition, the LEP electroweak working group has made available
a new set of precision electroweak measurements \cite{EW95} based
on increased statistics of over  $1.1\times10^7$ $Z$ decays. The
most significant  improvements have been a 50\% reduction in the
error on the $Z$ mass measurement, $M_Z=91.1887\pm0.0022$ GeV, and
a 30\% reduction in the error in the hadronic cross-section
measurement $\sigma^0_h=41.492\pm0.081$~nb. There have also been
significant reductions in the errors on $\Gamma_Z$, $A^\ell_{FB}$,
$A_\tau$ and $A_e$. Also important is a shift by more than one
standard deviation in the central value of $A^b_{FB}$ to
$0.1015\pm0.0036$, which brings it into significantly better
agreement with the global electroweak fit parameters. On the other
hand, the apparent discrepancy in $R_b$ has not been reduced. In
our analysis we assume that this apparent discrepancy is not due to
new physics.

%%%%%%%%%%%%%%%%%%%%%%%%%%%%%%%%%%%%%%%%%%%%%%%%%%%%%%%%%%%%%%%%%%%
%
%  Sommario relativo a ALR:
%
%  1991-1993: ALR = 0.1656+-0.0071(stat)+-0.0028(sys) Ref.LR94
%  1994-1995: ALR = 0.1524+-0.0042(stat)+-0.0012(sys) Ref.LR95
%  1991-1995: ALR = 0.1551+-0.0040(tot)               Ref.LR95
%
%  Corrispondenti valori di sin^2_eff:
%
%  1991-1993: sin = 0.2294+-0.0010                    Ref.LR94
%  1994-1995: sin = 0.2310+-0.0005                    Ref.LR95
%  1991-1995: sin = 0.2305+-0.0005                    Ref.LR95
%
%  Valore di sin2eff misurato a LEP:
%
%    1995   : sin = 0.2320+-0.0004                    Ref.EW95
%
%%%%%%%%%%%%%%%%%%%%%%%%%%%%%%%%%%%%%%%%%%%%%%%%%%%%%%%%%%%%%%%%%%

Concerning the SLD measurement of $A_{LR}$ (or, equivalently, of
$\sin^2\theta^{lept}_{eff}$), the latest value is \cite{LR95}:
$A_{LR}=0.1551\pm0.0040$
$(\sin^2\theta^{lept}_{eff}=0.2305\pm0.0005)$, corresponding to the
global 1991-95 SLD data sample. This should be compared to the
1991-93 SLD value \cite{LR94}: $A_{LR}=0.1656\pm0.0076$
$(\sin^2\theta^{lept}_{eff}=0.2294\pm0.0010)$, and to the LEP value
\cite{EW95}: $\sin^2\theta^{lept}_{eff}=0.2320\pm0.0004$. It is
evident that, as far as $\sin^2\theta^{lept}_{eff}$ is concerned,
the latest SLD  central value is now closer to the LEP value,
although the reduction of the SLD error means that the values are
still about $2\sigma$ apart. Nevertheless, we include $A_{LR}$ in
our global fit.

Other new elements in our fit are an updated value for the $W$ mass:
$M_W=80.33\pm 0.17$ GeV (world average) \cite{MW95}, and two new
measurements of parity violation in atomic Thallium that have
recently been reported: ${\cal
R}=Im\,\{E1^{PNC}/M1\}=(-15.68\pm0.45)\times10^{-8}$ \cite{PV01} and
$(-14.68\pm0.17)\times10^{-8}$ \cite{PV02}. The power of these two
Thallium experiments in constraining electroweak radiative
corrections is comparable to that of atomic Cesium results
\cite{PV03}. Apart from the inclusion of the above new atomic
result, our treatment of the available low-energy precision
electroweak data is identical with that documented in our previous
works \cite{El94,El92,El90}.  We emphasize that treating the
deep-inelastic $\nu N$ scattering cross-section measurements as
measurements of $1-M_W^2/M_Z^2$ (see, e.g., Ref.~\cite{EW95}) is
only an approximation, and that there are other significant
low-energy electroweak measurements that we include in the global
fit \cite{Char}.

Figure~1 shows the results of global fits within the Standard Model
to the available electroweak measurements, as contours of
$\Delta\chi^2=1,\,4$ in the $(M_H\,,\,m_t)$ plane. We recall that
their projections onto the coordinate axes correspond to
$\pm1\sigma$ and $\pm2\sigma$ errors on the top and Higgs masses.
The dashed lines are fits that do not include the combined CDF and
D0 measurements of $m_t$, which is shown as an error bar on the
left. Projecting the $\Delta\chi^2=1$ dashed ellipse on the
vertical axis, we find
\vskip4mm\noindent
%................................................................
\beq
    m_t = 156\begin{array}{l}+14\\ [-5mm]\\-15 \end{array}\quad\GeV
\eeq
%................................................................
\vskip4mm\noindent
for the Standard Model fit to the precision electroweak data with
$M_H$ left free, with a minimum $\chi^2_{\rm min}=12.2$. Our
central value of $m_t$ in (1) is somewhat lower than that quoted by
the LEP electroweak working group \cite{EW95}, mainly because we do
not fix the central value of $M_H$ at 300 GeV, and partly as a
result of our more complete treatment of the available low-energy
data (that prefer a relatively ``light'' top).  Fig.~2 shows the
contributions of the various different electroweak sectors to the
$\chi^2$ function of the global fit for the particular choice
$M_H=M_Z$, and Fig.~3 shows the global $\chi^2$ functions for a
sampling of different values of $M_H$. We have checked that, if we
restrict our fit to the LEP data alone, and assume the same value
of $M_H$ (300 GeV), our central value of $m_t$ agrees with theirs
within 3 GeV, which is within the typical theoretical uncertainties.

The small size of the error in (1) is a tribute to the precision of
the LEP experiments, in particular.  The range in (1) is compatible
with the CDF/D0 measurements, although somewhat lower. This
compatibility is  an impressive confirmation of the Standard Model
at the one-loop level, and justifies combining the direct and
indirect information on $m_t$. The solid lines in Fig.~1  are the
$\Delta\chi^2=1,\,4$ contours for such a combined fit, whose
projection on the vertical axis yields
\vskip4mm\noindent
%................................................................
\beq
           m_t=172\pm 10 \quad\GeV \quad.
\eeq
%................................................................
\vskip4mm\noindent
The $\chi^2_{\rm min}$ of the global fit is increased by
$\Delta\chi^2=1.8$  to  $\chi^2_{\rm min}=14.0$ when the CDF and D0
measurements of $m_t$ are included. This increase in $\chi^2$ is
acceptable $(<1.4\,\sigma)$, and the total $\chi^2$/d.o.f. remains
of order unity.

\section{Implications for $M_H$}

We now turn to the discussion of $M_H$, which is the main purpose of
this paper. Projecting the $\Delta\chi^2=1$ contours of Fig.~1 on
the horizontal axis, we find for the fit to the precision
electroweak data alone
\vskip4mm\noindent
%................................................................
\beq
M_H = 36\begin{array}{l}{+56}\\[-5mm]\\{-22}\end{array}
                                            \quad\GeV
    \qquad\left[\,\log_{10}(M_H/M_Z)=-0.40
        \begin{array}{l}{+0.40}\\[-5mm]\\{-0.41}\end{array}
                                                \right]\quad
\eeq
%................................................................
\vskip4mm\noindent
and for the fit that includes also the CDF/D0 $m_t$ measurement:
\vskip4mm\noindent
%................................................................
\beq
                   M_H = 76\begin{array}{l}{+152}\\[-5mm]\\
{-\,\,50}\end{array}\quad\GeV
     \qquad\left[\,\log_{10}(M_H/M_Z)=-0.08
\begin{array}{l}{+0.48}\\[-5mm]\\{-0.46}\end{array}\right]\quad.
\eeq
%................................................................
\vskip4mm\noindent
In each case, we have restated the fit result in a logarithmic
scale, since the leading dependences of the experimental observables
on $M_H$ are logarithmic. We note that the errors are fairly
symmetric in this scale, reflecting the fact that the $\chi^2$
function is well-behaved in $\log_{10}(M_H/M_Z)$ around the
absolute minimum. This is seen in Fig.~4, which displays $\chi^2$
as a function of $M_H$ (on a logarithmic scale) for a sampling of
choices of $m_t$. The envelope of these $\chi^2$ functions is the
$\chi^2$ function for $M_H$ with $m_t$ free, corresponding to the
projection of Fig.~1 on the horizontal axis.

We have verified that the shape of the $\chi^2$ function we find is
similar to that found by the LEP electroweak working group
\cite{EW95} if we restrict our fit to a similar data set. It is
clear that the default value $M_H=300$ GeV assumed by the LEP
electroweak working group in quoting central values of $m_t$ is not
the most probable value, and is indeed more than $1\sigma$ away far
from it. We stress again that, because of the well-known positive
correlation between $m_t$ and $M_H$ visible in Fig.~1, this assumed
value of $M_H$ is the main reason the  LEP electroweak working
group quotes a higher central value of $m_t$ than we do in (1) and
(2).

We have also verified that the shape of the $\chi^2$ function found
by Swartz \cite{Swapc} in a fit using a very similar data set is
similar to ours, though obtained with a different treatment of the
low-energy data.

\section{Variations in the Analysis}

Before discussing the predicted range [eqs. (3), (4)] of $M_H$  in
more detail, we comment on how our analysis is affected by
uncertainties in $\alpha_s(M_Z)$ and by the $A_{LR}$ measurement at
SLD. Then we consider in more detail the impact of $\alpha(M_Z)$.
In Fig.~1, $\alpha_s(M_Z)$ is fixed at the best fit value, 0.124
(to which we attach an uncertainty of $\pm0.005$). This is somewhat
higher than the world average: $\alpha_s(M_Z)=0.117\pm0.007$
\cite{Be94}. Imposing $\alpha_s(M_Z)=0.117$ in the fit, the value
of $\chi^2_{min}$ increases by $\sim\!1.8$, but $M_H$ diminishes by
only $\sim\!7$ GeV, and $m_t$ is not significantly affected. We
conclude that the uncertainty in $\alpha_s(M_Z)$ is not an
important factor at present in the analysis of $M_H$.

Concerning $A_{LR}$, it is well-known that the SLD value tends to
bring $M_H$ down with respect to the rest of the electroweak data
\cite{El94}. However, even excluding $A_{LR}$ completely, a
procedure that we do not consider justified, we find that the
central values of $\log_{10}(M_H/M_Z)$ in eqs. (3) and (4) are
increased  by about +0.16 and +0.28 respectively, namely less than
the corresponding $1\sigma$ uncertainty in $\log_{10}(M_H/M_Z)$.

We now turn to the sensitivity of our results to the assumed value
of $\alpha(M_Z)$. In the past, we have taken
$\alpha(M_Z)^{-1}=128.87\pm0.12$ from Ref.~\cite{Je91}, but
recently there have been several re-evaluations of the
extrapolation from the Thompson limit,  some of which differ
appreciably from the earlier value \cite{Je91}. In this paper we
have assumed $\alpha(M_Z)^{-1}=128.896\pm0.090$ from
Ref.~\cite{Eide}, which is similar to the recent estimate in
Ref.~\cite{Burk} ($128.89\pm0.090$). We now   explore the
implications  of varying $\alpha(M_Z)^{-1}$
within the range suggested by other estimates \cite{Mart,Swar}
%___________
\footnote{
The central value in Ref.~\protect\cite{Mart} ($128.99\pm0.06$) and
Ref.~\protect\cite{Swar} ($129.08\pm0.10$) are about 1 and
$2\sigma$ above the central values in
Refs.~\protect\cite{Eide,Burk}. We have recently been informed [M.
Swartz, private communication] that an update of
Ref.~\protect\cite{Swar} yields a value close to
Refs.~\protect\cite{Eide}.  }.
%___________
Fig.~5 shows the values of a subset of electroweak observables
($\sin^2\theta_{eff}^{lep}$, $M_W$ and $\Gamma_Z$)  in the $m_t$
range indicated by CDF and D0 and for three choices of $M_H$ (this
is not a  fit).  The left-hand side of the figure is for
$\alpha(M_Z)^{-1}=128.896\pm0.090$, and the right-hand side for a
value $2\sigma$ higher, namely $\alpha(M_Z)^{-1}=129.076\pm0.090$,
similar to the evaluation of \cite{Swar}. The minor axes of
the theoretical ellipses in Fig.~5 are due to the propagation of
the error in $\alpha(M_Z)^{-1}$. We see that the effects on $M_W$
and $\Gamma_Z$ of varying $\alpha(M_Z)$ are very small, and that
the effect on $\sin^2\theta_{eff}^{lep}$ is to  bring the
theoretical predictions closer to the SLD measurement. However, it
is evident in the same Fig.~5 that even this $2\sigma$ shift in the
electromagnetic coupling constant is less relevant in the
theory/experiment comparison than the dispersion of the LEP/SLD
data, and thus it does not affect significantly the stability of the
$M_H$ range discussed previously.

Conversely, we can ask if future precision electroweak data can
improve our knowledge of $\alpha(M_Z)$. A $2\sigma$ variation in
$\alpha(M_Z)^{-1}$ can induce a few GeV shift in $m_t$ at fixed
$M_H$ (see, e.g., Ref.~\cite{Swar}), so it is not impossible that
the combination of future, more precise direct (CDF/D0) and
indirect (LEP/SLD) determinations of $m_t$ with an error of
$\sim\!5$ GeV could also reduce implicitly the uncertainty in
$\alpha(M_Z)^{-1}$.

Such future improvements may also be linked to future more precise
$g_\mu\!-\!2$ measurements possible with the  BNL E821 experiment
\cite{E821}. The reason is that the theoretical determinations of
the hadronic contribution to $\alpha(M_Z)^{-1}$ and $g_\mu\!-\!2$
are correlated, since the same set of $e^+e^-\rightarrow hadrons$
data is used in their dispersion integral estimates, although with
different convolution kernels. Assuming full correlation of the
partial systematic errors induced in $\alpha(M_Z)$ and
$g_{\mu}\!-\!2$ by the different independent low-energy subsets of
the data compiled in Ref.~\cite{Eide}, we have estimated the
theoretical joint standard deviation ellipse in the
$[\alpha(M_Z),\,g_{\mu}\!-\!2]$ plane (Fig.~6).  Also shown in
Fig.~6 is the situation (dotted ellipse) to be expected after
prospective improvements in measuring hadron production at
DA$\Phi$NE and VEPP-2M (see \cite{Eide} and references therein),
where we see that the correlation between $\alpha_{em}(M_Z)^{-1}$
and $g_\mu\!-\!2$ becomes stronger. Also shown is the prospective
error in $g_\mu\!-\!2$ expected to be obtained by the BNL E821
experiment (horizontal band). We see that, by virtue of this
correlation, the anticipated measurement in this experiment could
even serve to constrain the possibile range of $\alpha(M_Z)$. The
vertical band reminds us the possibility of fitting a value of
$\alpha(M_Z)$ from future precision data, as is now done with
$\alpha_s(M_Z)$, though its width is purely hypothetical.

\section{Implications of Our Analysis of $M_H$}

In view of the remarkable stability of the $M_H$ range in Fig.~1,
the indication for a relatively light Higgs mass of ${\cal O}(M_Z)$
should be taken seriously. The upper limit at $2\sigma$
($M_H\lsim700$ GeV, including CDF/D0) is reassuringly below the TeV
region, so the perturbative calculations which the fit is based
upon are expected to be reliable. The upper end of the $1\sigma$
range ($M_H\lsim230$ GeV) and the central value $M_H=76$ GeV give
hope for finding the Higgs at the LEP2 or the LHC.  In general, it
is definitely non-trivial that the electroweak data consistently
favour a Higgs mass in a range of ${\cal O}(M_Z)$, which disfavours
composite or strongly-interacting scenaria, as discussed elsewhere
\cite{El95}.

The question arises whether this range is compatible with bounds on
the SM Higgs mass derived from (meta)stability of the electroweak
vacuum, and from perturbative behaviour of the SM couplings. In the
upper part of Fig.~7 we plot first the same $\Delta\chi^2=1$
contour as in Fig.~1 (CDF/D0 included), the dashed part
representing the LEP direct limit $M_H>65$ GeV. Superposed are the
lower limits on $M_H$ from vacuum metastability requirements
\cite{Meta}, as a function of the ``new physics'' scale $\Lambda_V$
in GeV up to which the effective potential in the SM is assumed to
apply (bounds from absolute stability of the SM vacuum \cite{Stab}
would be weaker by a few GeV for our central value of $m_t=172$
GeV). The $M_H$ range we find is compatible with the
(meta)stability bounds, particularly if $\Lambda_V$ is small, but it
is not yet possible  to exclude any value of $\Lambda_V$ and thus
give any indication on the possible new physics scale. Also shown
in Fig.~7 are upper bounds on $M_H$ obtained by requiring the SM
couplings to remain perturbative up to a scale $\Lambda_P$. We see
that these are also compatible with our analysis, particularly if
$\Lambda_P$ is small, though again we cannot exclude any range of
this scale. In the particular case $\Lambda_V=\Lambda_P=10^{19}$
GeV and $m_t=172$ GeV (our central value), these bounds become 116
GeV $< M_H <$ 190 GeV.

In the minimal supersymmetric extension of the Standard Model
(MSSM) the Higgs sector depends on the pseudoscalar Higgs mass
$M_A$, the v.e.v. ratio $v_2/v_1=\tan\beta$ and the value of the
top mass, through radiative corrections to the Higgs potential. We
assume fixed, large values for the other MSSM parameters, so that
the remaining MSSM spectrum decouples. Then, for any given value of
$\tan\beta$, the radiative corrections induced by the MSSM Higgs
sector are specified by the lightest Higgs mass $m_h$ and $m_t$,
which are the coordinates of the lower plot in Fig.~7. The previous
metastability bounds do not apply to the MSSM vacuum. However, new
intrinsic {\em upper\/} bounds on $m_h$ appear, as shown for two
representative values of $\tan\beta$ ($\tan\beta=2,\,16$). For
$m_t=172$ GeV (our central value), the upper limit on $m_h$ in the
MSSM is $124$ GeV. For $m_h\sim{\cal O}(M_Z)$, the radiative
corrections arising from the SM and MSSM Higgs sector differ only
by small subleading terms, and the similarity of the $\chi^2$
functions in the SM and the MSSM has been demonstrated in previous
analyses  \cite{El94,El95}, hence the similarity of the
$\Delta\chi^2 =1$ contours in the upper and lower halves of Fig.~7.

We conclude this paper by proposing an exploratory interpretation
of our results addressed to a possible comparison between the SM
and the MSSM. In Fig.~8 we show the cumulative probabity $P(M_H)$,
calculated from the behaviour of the SM $\chi^2$ function shown in
Fig.~4, integrated appropriately over $m_t$ and including the
measurements from CDF \cite{CD95} and D0 \cite{D095}. We note
that this full cumulative probability distribution does not apply to
the MSSM, because of the intrinsic upper limit on $m_h$ mentioned
in the previous paragraph. However, we can use the cumulative
probability curve in Fig.~8 to compare the SM and the MSSM by
estimating the relative probabilities of the mass ranges allowed in
the two models when other experimental and/or theoretical
constraints, not incorporated in the structure of the electroweak
radiative corrections, are taken into account. This comparison may
be made using the SM curve in Fig.~8, because, as already mentioned,
the $\chi^2$ functions for the SM and the MSSM are quite similar in
the mass range around $M_Z$ which contains the bulk of the
probability distribution \cite{El94}.

In the case of the SM, we have a direct experimental lower limit
$M_H > 65$ GeV \cite{EW95}, but also the stronger metastability
lower bound of $116$ GeV and the perturbative upper bound of $190$
GeV mentioned earlier. We estimate from Fig.~8 a total probability
of $18 \%$ for the mass range $116$ GeV $< M_H < 190$ GeV. In the
case of the MSSM, the direct experimental lower bound on $m_h$ is
somewhat weaker, and may be taken as $50$ GeV, and there is no
metastability lower bound, only the intrinsic upper bound of $124$
GeV. We estimate from Fig.~8 a total probability of $36 \%$ for the
mass range $50$ GeV $< M_H < 124$ GeV. The relative probability is
clearly higher for the MSSM than for the SM, but not significantly
so.

The limitations and approximations inherent in this exploratory
analysis are many and obvious. However, it provides us with a clear
message: the data are surprisingly consistent with the MSSM,
perhaps even more consistent than with the SM.
\pagebreak

%\vspace*{5mm}
\noindent{\Large\bf Acknowledgements}
\vspace*{3mm}

The work of E.L.\ is supported in part by an I.N.F.N.\ post-doctoral
fellowship, and by funds of the Institute of Advanced Study
(Princeton).

\newpage
\noindent{\Large\bf Figure Captions}
\vspace*{3mm}
\begin{description}

\item{Fig.~1 --} Combined fit to all precision electroweak data
                 in the $(M_H,\,m_t)$ plane, including (solid lines)
                               or not (dashed lines) the direct determination
                               of $m_t$ by CDF/D0 (error bar on the left).
                               The contours correspond to $\Delta\chi^2=1,\,4$
                               around the minimum (small circle) in either
case.
                         Notice that $M_H$ is significantly below 300 GeV
                 at the $1\sigma$ level, and below 1 TeV at the
                 $2\sigma$ level.

\item{Fig.~2 --} The contributions to $\Delta\chi^2$ due to
                 different sectors of the precision electroweak data
                 set, as functions of $m_t$ for an assumed value
                 $M_H = M_Z$.

\item{Fig.~3 --} The values of $\chi^2$ as functions of $m_t$ for
                 the various indicated values of $M_H$.

\item{Fig.~4 --} The values of $\chi^2$ as functions of $M_H$ for
                 the various indicated values of $m_t$.

\item{Fig.~5 --} The impact of a hypothetical shift of
                 $\alpha(M_Z)^{-1}$
                 on selected electroweak observables
                               $(\sin^2\theta^{lept}_{eff},\,M_W,\,\Gamma_Z)$.
The
                 three sub-figures on the left show the predictions
                 (slanted ellipses) for such observables, assuming:
                 the indicated value of $\alpha(M_Z)^{-1}$ (error
                 included) \cite{Eide}, the CDF/D0 measurement of
                 $m_t$, and three representative values of $M_H$
                 (65, 300 and 1000 GeV). The gray horizontal stripes
                 represent the corresponding experimental
                 determinations. If the central value of
                 $\alpha(M_Z)^{-1}$ is increased by 2 standard
                 deviations, the three sets of predictions on the
                 right are obtained.
                 Notice that the most significant effect is that on
                 $\sin^2\theta_{eff}^{lept}$.

\item{Fig.~6 --} One-standard-deviation ellipse corresponding
                 to present theoretical estimates \cite{Eide} of
                 $\alpha(M_Z)^{-1}$ and $g_\mu-2$ (solid line).
                 Also shown as a dotted ellipse is the envisaged
                 reduction in the uncertainty that will come from
                 future low-energy experiments (mainly DA$\Phi$NE).
                 Notice the non-negligible correlation in both
                 cases. The gray horizontal stripe represents a
                 possible outcome of the high-precision
                 $g_\mu\!-2\!$ experiment E821 at BNL \cite{E821}.
                 The gray vertical stripe reminds us that some
                 valuable indirect information on $\alpha(M_Z)^{-1}$
                 will be provided by the combination of more precise
                 future electroweak measurements.

\item{Fig.~7 --} Comparison of combined top-Higgs mass fits in the
                 Standard Model (SM, upper plot) and in its Minimal
                 Supersymmetric extension (MSSM, lower plot), at
                               $\Delta\chi^2=1$. The continuation of the
                 $\Delta\chi^2 = 1$ contour below the LEP direct
                 limit $M_H>65$ GeV is shown dashed. Also shown
                         in the SM plot are the lower limits on $M_H$ from
                 vacuum metastability \cite{Meta}
                 as a function of the ``new physics''
                        scale $\Lambda_V=10^4$--$10^{19}$ GeV,
                 and the upper limts that come from requiring the SM
                 couplings to remain perturbative up to a scale
                 $\Lambda_P=10^3$--$10^{19}$ GeV. In the MSSM plot,
                 the dashed region indicates $m_h$ below $50$ GeV, and
                 we show the intrinsic upper limits on the lightest
                 Higgs mass for two values (2 and 16) of
                 $\tan\beta=v_2/v_1$.

\item{Fig.~8 --} The cumulative probability distribution calculated
                 from the $\chi^2$ function in the SM shown in
                 Fig.~4, obtained after integrating
                 appropriately over $m_t$, including the direct
                 measurements from CDF \cite{CD95} and D0
                 \cite{D095}. This may be used to estimate the
                 relative probabilities of different Higgs mass
                 ranges in the SM and the MSSM, as discussed in
                 the text.

\end{description}

\end{document}